\documentclass[runningheads]{llncs}
\usepackage{amsmath,amssymb}
\usepackage{graphicx}
\usepackage[vlined, ruled, linesnumbered]{algorithm2e}
\begin{document}
\title{Zealotry and Influence Maximization in the Voter Model: When to Target Zealots?}
\titlerunning{Zealotry and Influence Maximization in the Voter Model}
%
\author{Guillermo {Romero Moreno}*\orcidID{0000-0002-0316-8306} \and\\
Edoardo Manino\orcidID{0000-0003-0028-5440} \and
Long Tran-Thanh\orcidID{0000-0003-1617-8316} \and
Markus Brede\orcidID{0000-0002-0197-8612}}
\authorrunning{G. Romero Moreno et al.}
%
\institute{School of Electronics and Computer Science,\\
University of Southampton, Southampton, UK\\
\email{*grm1g17@soton.ac.uk}}
\maketitle              
\begin{abstract} 

In this paper, we study influence maximization in the voter model in the presence of biased voters (or zealots) on complex networks. Under what conditions should an external controller with finite budget who aims at maximizing its influence over the system target zealots? Our analysis, based on both analytical and numerical results, shows a rich diagram of preferences and degree-dependencies of allocations to zealots and normal agents varying with the budget. We find that when we have a large budget or for low levels of zealotry, optimal strategies should give larger allocations to zealots and allocations are positively correlated with node degree. In contrast, for low budgets or highly-biased zealots, optimal strategies give higher allocations to normal agents, with some residual allocations to zealots, and allocations to both types of agents decrease with node degree. Our results emphasize that heterogeneity in agent properties strongly affects strategies for influence maximization on heterogeneous networks.

\end{abstract}
\section{Introduction}
Perhaps motivated by the increasing prevalence of social media and their influence on public opinion, processes of opinion formation on social networks have found much attention in the recent literature~\cite{Acemoglu2011,Chen2013}. Models in this domain have addressed general properties of opinion diffusion on static and co-evolving networks (refer to, e.g., \cite{Castellano2009,Sirbu2017} for reviews), but also basic mechanisms underlying phenomena such as radicalization \cite{Ramos:2015} and the role of external influence~\cite{De2018,Palombi2017}.

Whilst typical models of opinion formation often include the role of external influence, 
another relevant aspect is the question of its placing, a problem that has been termed \emph{influence maximization} (\emph{IM}). Influence maximization has huge relevance for a variety of important applications that range from advertising \cite{Domingos2001}, political and public information campaigns \cite{WilderHIV,Zhang2016}, to questions on how to optimally encourage a developing economy \cite{Alshamsi2018}. Starting with \cite{Kempe2003}, the above problem has a huge pedigree in the computer science literature and has traditionally been studied in the context of the independent cascade (IC) model \cite{Goldenberg2001} or variants thereof such as threshold models \cite{Granovetter1978}. In the typical problem setting for IC-like models, one seeks optimal seed nodes from which a one-off cascade can reach as large  as possible a part of the social network.

More recent examples of strategically exerted interference on social media aiming at disrupting the democratic process~\cite{Russia:techrep,Badawy2018} have sparked the interest of influence maximization on election events~\cite{Wilder:2018:CET:3237383.3237428}. 
Here, we want a mechanism of opinion formation in which the agents can repeatedly change their opinion. In contrast to the typical IC literature, where flips of opinion only occur in one direction, these models are more suitable for volatile opinions or longer time frames, as these conditions fit better the political domain.
In the following, we focus on the voter model (VM)~\cite{Clifford1973,Holley1975} because of its simplicity and prominence in the literature. A key feature of the VM is its linearity, which makes it possible to establish many of its properties~\cite{Redner2019}.
Previous work on the VM has also considered aspects of agent heterogeneity. For this purpose, existing literature has introduced so-called zealots, i.e. agents that have decreased chances of adopting particular opinions. The VM has been studied with perfect zealots whose opinions are completely unaffected by other individuals~\cite{Mobilia:2003,Kuhlman2013,Masuda2015} (also referred to as ``inflexible'', ``committed'', ``stubborn'' or ``frozen'') and imperfect zealots \cite{Masuda2010} (also ``partisans''), who can change their opinion albeit changes are biased towards one of the opinions. Dynamics including zealots from different opinions typically result in mixed equilibrium states. 

Previous work on influence maximization in the VM initially sought the initial opinion distribution in the network that would lead to the preferred opinion in the consensus state with the highest probability~\cite{Even-Dar2011}. However, in the presence of zealots equilibrium states of the VM are independent of initial opinion distributions, so later works shifted the goal to optimally transforming nodes into zealots to reach improved mixed-equilibrium states~\cite{Kuhlman2013,Yildiz2014}. Another approach, more akin to problems in network control~\cite{Liu2011,Porfiri2008}, regards zealots as external controllers who optimally build unidirectional influence links (edges) to nodes in the network~\cite{Masuda2015,Brede2018,BredeHor2018}. In this approach, external controllers can be seen as political parties, mass media or advertising companies that must choose the targets of their campaign policies.

As pointed out by Aral and Dhillon~\cite{Aral:2018}, past approaches to influence maximization often overlook aspects of heterogeneity in agent behavior. Specifically, in the context of the IC, it has been shown that results of IM are strongly affected by agents' susceptibility to adopt opinions~\cite{Aral:2018} ---or, translated into the context of the VM, zealotry. However, whereas demonstrating that zealotry influences optimal allocations in the IM, Aral and Dhillon~\cite{Aral:2018} have not explored detailed mechanisms for such differences. Here, we extend previous work on IM for the voter model with external controllers by considering the effects of different levels of imperfect zealotry in the population. An optimal campaign manager with limited resources would never target perfect zealots whose opinions cannot be influenced. However, under what conditions would she target imperfect zealots and how does this relate to the zealot's topological position in the social network? Below, we shall explore these questions for different network topologies. As zealots represent partly radicalized agents, answers to the questions might help to find ways of reducing radicalism in social systems.

Our paper is organized as follows. Section \ref{sect:model} formulates the proposed IM problem and introduces solution methods. Section \ref{sect:results} first presents analytical and numerical solutions to the problem on simple graph topologies and later extends the analysis to scale-free networks. Section~\ref{sect:conclusions} discusses and summarizes our main conclusions.

\section{Model and methods} \label{sect:model}
We introduce a scenario where an external controller wants to influence a group of $N$ individuals connected via a social network. We assume individuals hold binary opinions $o_i\in\{A,B\}, i=1,\ldots,N$ and define $x_i$ as the probability of node~$i$ holding opinion $A$. Opinion diffusion occurs following the usual VM, where a random node copies the opinion of a random neighbor at each time-step.
We assume nodes have a bias against opinion $A$, which we model by an intrinsic level of zealotry $q_i\in[0,1]$ that gives the probability of not copying opinion $A$ from a neighbor~$j$ who holds $o_j=A$. Below, we are interested in optimal strategies for an external controller who wants to promote opinion $A$ in the network. As in \cite{Masuda2015,Brede2018,BredeHor2018}, a controller is modeled as a node with fixed opinion that exerts influence by establishing unidirectional links to nodes in the network with a specific link weight $a_i>0$. Overall, the controller wants to optimize the allocation of the $a_i$ to maximize her vote share, subject to a budget constraint $\mathcal{B}_a\geq\sum_i a_i$. Note that, unlike previous work, we relax the conditions on the $a_i$ by allowing them to adopt any value within $\mathbb{R}$.

To proceed, we write rate equations for the probabilities of nodes holding opinion $A$, which evolve according to
\begin{equation}
    \frac{dx_i}{dt} = (1-q_i) \:(1-x_i)\frac{\sum_j w_{ij} x_j + a_i}{k_i + a_i}- x_i\frac{\sum_jw_{ij}(1-x_j)}{k_i+a_i} \; , \label{eq:rate_eq}
\end{equation}
where $W=(w_{ij})$ is the weighted adjacency matrix and $k_i=\sum_jw_{ij}$ is the weighted in-degree of node $i$. Note that the left term on the right-hand side of the equation accounts for the probability of a node holding opinion $B$ and copying the opinion of a neighbor with $A$ (which is reduced by the level of zealotry $q_i$), while the right term reflects the opposite situation.

We focus on influence maximization in the expected equilibrium state, which is unique and asymptotically reached irrespective of initial conditions \cite{Yildiz2014}. The probabilities of adopting $A$ in equilibrium, $x_i^*$, can be determined from $dx_i/dt=~0$, leading to the system of $N$ second-order equations
\begin{equation}
    0 = (1-q_i) \:\sum_j w_{ij} x_j^* + q_i\:x_i^*\sum_jw_{ij}x_j^* - x_i^*\:(k_i+(1-q_i)\:a_i) + (1-q_i)\:a_i \; . \label{eq:equilibrium}
\end{equation}
Since this system of equations is hard to solve numerically, the vote shares in the equilibrium state can alternatively be found by performing numerical integration of \eqref{eq:rate_eq}.

The external controller aims to distribute her unidirectional connections in such a way as to maximize the expected vote share in equilibrium $X^*=\frac{1}{N}\sum_i x_i^*$, leading to the optimization problem
\begin{equation}
    \max_{\boldsymbol{a}} X^*(W,\boldsymbol{a},\boldsymbol{q})  \qquad s.t.\quad \mathcal{B}_a\geq\sum_i a_i\;,\quad a_i\geq0\;.
\end{equation}
As we lack an explicit expression for the total vote share $X^*$ on general networks, we resort to numerical methods for the optimization process. The continuous definition of influence allocations allows gradient ascent techniques~\cite{Unpublished}, for which we need to compute the gradient of the equilibrium vote share with respect to the allocations, $\nabla_{\boldsymbol{a}} X^*$. By applying partial derivatives on \eqref{eq:equilibrium}, we obtain the gradients as
\begin{multline*}
    0 = (1-q_i) \:\sum_j w_{ij} \frac{\partial x_j^*}{\partial a_l}
    + q_i\left( \frac{\partial x_i^*}{\partial a_l}\sum_jw_{ij}x_j^* + x_i^*\sum_jw_{ij}\frac{\partial x_j^*}{\partial a_l}\right) -\\
    -\frac{\partial x_i^*}{\partial a_l}[k_i + (1-q_i)\:a_i] + \delta_{i,l}(1-q_i) (1 - x_i^*)\;,
\end{multline*}
\begin{multline}
    \nabla_{\boldsymbol{a}} X^* = \frac{1}{N} (\nabla_{\boldsymbol{a}} \boldsymbol{x}^*)^T \boldsymbol{1} = \frac{1}{N}\;\mathrm{diag}\left[(1-q_i)(1-x_i^*)\right] \; [\mathrm{diag}(k_i + (1-q_i)\:a_i) -\\
    - \mathrm{diag}\left(q_i\right)\mathrm{diag}(W\boldsymbol{x}^*) - W\mathrm{diag}(1-q_i + q_i x_i^*)]^{-1}\; \boldsymbol{1} \; , \label{eq:gradient}
\end{multline}
where $\delta_{i,l}$ is the Kronecker delta and the values of $\boldsymbol{x}^*$ are computed via numerical integration, as explained above.

\section{Results} \label{sect:results}
To gain intuition about the role of zealots in influence maximization in the VM and have some analytical reference, we first explore simple graph topologies in Sec. \ref{subsect:simple}. We then extend the experiments to networks with heterogeneous degree distributions in Sec. \ref{subsect:sf}, as their network structure is closer to the ones found in real social network data.

\subsection{Simple Network Topologies} \label{subsect:simple}
\subsubsection{Complete Graph.}
We start with an infinite complete graph where a fraction $\rho$ of nodes are zealots with equal level of zealotry $q_z$ and the rest of the nodes are normal (unbiased) agents with $q_n=0$.
We assume that the external controller allocates the same link strength to all nodes within a group, where $a_z$ and $a_n$ are the weights of link allocations to zealots and normal agents as a fraction of nodes in the network. Inserting the budget constraint, we find
\begin{equation*}
\mathcal{B}_a / N^2 = \langle a\rangle = \rho\,a_z + (1-\rho)\,a_n=\alpha\langle a\rangle+(1-\alpha)\langle a\rangle \; ,
\end{equation*}
where $\alpha\in[0,1]$ is the unique decision parameter for the distribution of the budget among zealots and normal agents. The whole influence budget is focused on normal agents when $\alpha=0$, on zealots when $\alpha=1$, and link weights are equal for agents in both groups when $\alpha=\rho$.

The total vote share in the equilibrium for these settings can easily be derived from the rate equation in \eqref{eq:rate_eq}, leading to
\begin{equation}
    X^* = \rho\,x_z^* + (1-\rho)\,x_n^* =
    \frac{\langle a\rangle}{q_z\rho} \,
    \frac{(\alpha-\alpha\,q_z)\langle a\rangle(1-\alpha) + (1-\rho)\rho (1 - \alpha\,q_z)} {\langle a\rangle(1-\alpha) + (1-\rho)\rho} \; , \label{eq:complete_X}
\end{equation}
with a boundary at $X^*\!=1$. The vote share generally increases with budget availability $\langle a\rangle$ and decreases with the fraction of zealots $\rho$ and their level of zealotry $q_z$. Optimal allocations $\alpha^*$ can be found by solving $\partial X^*\!/\partial \alpha=0$, giving
\begin{equation}
    \alpha^*= 1 - \frac{\rho(1-\rho)}{\langle a\rangle} \left(\frac{1}{\sqrt{1-q_z}} - 1\right) \; , \label{eq:opt_alpha}
\end{equation}
which is bounded at $\alpha^*\!=0$. This result shows that, for a given share of zealots~$\rho$, optimal allocations target zealots the more the bigger the budget $\langle a\rangle$ and the smaller the level of zealotry $q_z$. We can find the switching point of zealotry $q_z^*$ at which normal agents start to be favored over zealots, corresponding to
\begin{equation}
    q^*_z = 1 - \left(\frac{1}{\langle a\rangle/\rho+1}\right)^2 \; , \label{eq:switching_q}
\end{equation}
which increases with the budget $\langle a\rangle$ and decreases with the density of zealots $\rho$.

\begin{figure}
    \includegraphics[width=\textwidth]{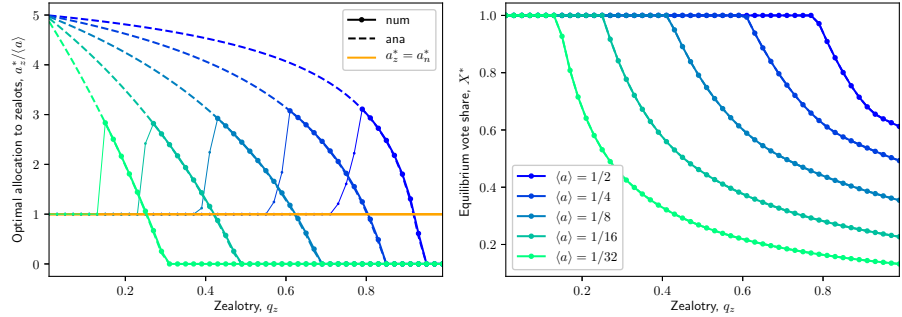}
    \caption{Normalized link allocations given to zealots (left) and resulting equilibrium vote shares (right) obtained via analytical (\emph{dashed}) and numerical (\emph{solid)} methods on a complete network for varying zealotry $q_z$ and varying budget $\langle a\rangle = \mathcal{B}_a/N^2$. The network has size $N=100$, with 20\% of zealots with zealotry $q_z$ and 80\% of normal agents with $q_n=0$. The horizontal line (\emph{orange}) represents equal link weights given to both groups. Numerical and analytical results for the vote share are in almost perfect agreement at all explored points.} \label{fig:complete}
\end{figure}

We use the analytical solutions developed above to evaluate the quality of numerical results obtained via gradient ascentas introduced in Sec.~\ref{sect:model}.
For this purpose, we analyze how (normalized) optimal allocations to zealots $a_z^*/\langle a\rangle=\alpha^*/\rho$ vary with the level of zealotry $q_z$ and budget size $\langle a\rangle$ for both methods (Fig.~\ref{fig:complete}--left) and observe the resulting equilibrium vote shares $X^*$ (Fig.~\ref{fig:complete}--right).
We note that for low values of zealotry $q_z$ or high budget $\langle a\rangle$, the controller achieves full control $X^*\!=1$ by both methods despite differences in their solutions of optimal strategy. Numerical results are initialized at targeting the two groups of agents equally ($a_z=a_n$) and deviate from this allocation in the direction of the gradient the minimum required to achieve $X^*\!=1$. Conversely, analytical optimizations do not take into account the boundary at $X^*\!=1$ and provide the most robust strategy.
For higher values of $q_z$ or lower budgets $\langle a\rangle$ ---where full control is not possible--- allocation strategies from both methods are found to be in perfect agreement, as well as the equilibrium vote shares achieved by them.

In general, optimal strategies favor allocations to zealots over allocations to normal agents for low values of $q_z$ and high values of $\langle a\rangle$. This behavior gradually decreases with $q_z$, switching to favoring normal agents at some $q_z^*$ (in the figure, crossing the horizontal orange line) and fully avoiding zealots ($a^*_z=0$) at another critical level of zealotry $q_z^{**}$. Both critical points depend on the available budget~$\langle a\rangle$. Equilibrium vote shares reach full control $X^*=1$ for low values of $q_z$, experience a steep drop just after leaving the full control regime and gradually decelerate in their decrease as $q_z$ increases.

\subsubsection{Bipartite Graphs.}
Next, as a way to explore the interplay of degree heterogeneity with zealotry, we explore complete bipartite graphs, i.e. graphs divided into two groups where all nodes from a group are connected to all nodes in the other group. We assume that nodes in the smaller group (\emph{hubs}) comprise a fraction $\rho<0.5$ of the total network and have zealotry $q_h$, while nodes from the larger group (\emph{periphery}) have zealotry $q_p=1-q_h$. The equilibrium vote share for this case is given by
\begin{equation}
    X^*\!=
    \frac{\langle a\rangle}{q_h\rho}
    \frac{(\alpha\!-\alpha q_h)\overline{\alpha}\langle a\rangle\!+ \overline{\rho}\rho (1\!-\! \alpha\,q_h)}
    {\overline{\alpha}\langle a\rangle + \overline{\rho}\rho(1+\overline{q_h}/q_h^2)}
    \!+\!
    \frac{\langle a\rangle}{\overline{q_h}\,\overline{\rho}}
    \frac{(\overline{\alpha} - \overline{\alpha}\,\overline{q_h})\alpha\langle a\rangle + \rho\overline{\rho}(1\!-\!\overline{\alpha}\,\overline{q_h})}
    {\alpha\langle a\rangle + \rho\overline{\rho}(1+q_h/\overline{q_h}^2)} \; , \label{eq:bipartie_X}
\end{equation}
where, for convenience of notation, overlined symbols mean $\overline{\gamma}=1-\gamma$ and there is again a boundary at $X^*\!=1$. Optimal allocations $\alpha^*$ can again be found by solving $\partial X^*\!/\partial \alpha=0$, which results in a fourth-order polynomial equation that we evaluate numerically.

\begin{figure}
    \includegraphics[width=\textwidth]{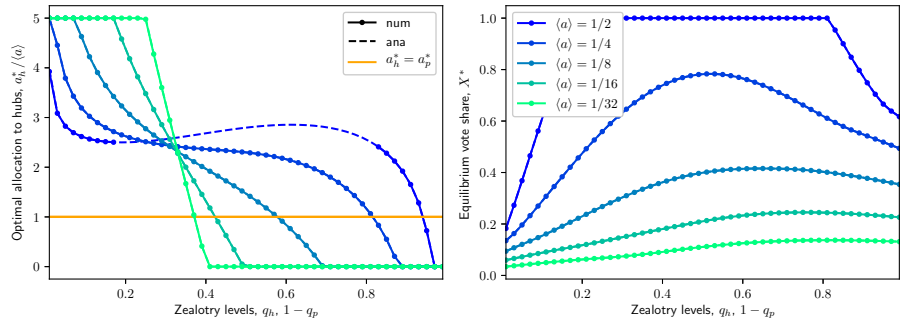}
    \caption{Normalized influence allocations given to hubs (left) and resulting equilibrium vote shares (right) for different levels of zealotry and budget availability $\langle a\rangle$ obtained via analytical (\emph{dashed}) and numerical (\emph{solid)} methods on a bipartite network of size $N=100$. Hubs, with zealotry $q_h$, compose 20\% of the network, while periphery nodes compose the remaining 80\% and have zealotry $q_p=1-q_h$. Optimal influence allocations from numerical methods are omitted when $X^*=1$ for improved clarity of the graph. Resulting equilibrium vote shares from numerical and analytical approaches match almost perfectly (see overlapping lines).} \label{fig:bipartite}
\end{figure}

Figure \ref{fig:bipartite} analyzes optimal allocations given to zealots on a bipartite network for the level of zealotry defined above, obtained with both analytical and numerical methods.
Figure~\ref{fig:bipartite}--left gives (normalized) optimal link allocations to hubs $a_h^*/\langle a\rangle=\alpha^*/\rho$ for different levels of zealotry $q_h$ and budget size $\langle a\rangle$. We note that, for a large budget (blue line), there is a consistent preference in allocations toward hub nodes regardless of which group holds a higher level of zealotry ---except in the limit of hubs with almost-perfect  zealotry ($q_h\approx1$). On the contrary, for a low budget (green line), optimal allocations quickly switch from fully targeting hubs to fully targeting periphery nodes, even when the level of zealotry of hubs is larger than that of periphery nodes ($q_h<q_p$).
Figure~\ref{fig:bipartite}--right gives the resulting equilibrium vote shares $X^*$ for optimal allocations. We note that, for low budgets (green lines), the vote shares obtained in optimal allocations are higher when zealots are concentrated on hubs. For high budgets (blue lines), moderate levels of zealotry in both groups ($q_z\approx q_h\approx0.5$) lead to higher vote shares from optimal allocations.

To summarize our findings on simple network topologies, we have seen that influence  allocations to zealots are preferred when their level of zealotry is low or the available budget is high, while allocations to normal agents are preferred otherwise. When nodes of different degrees are present, allocations to hubs are preferred over periphery nodes when the budget availability is high, even when the level of zealotry of hubs is higher than that of periphery nodes. In contrast, for low budgets, higher influence allocations to periphery nodes can be preferable in cases where their level of zealotry is higher than that of hubs.

\subsection{Scale-Free Networks. Zealotry and Node Degree} \label{subsect:sf}
Next, we focus on Barabasi-Albert (BA) networks \cite{Barabasi1999} to further explore how the interplay of zealotry and node degree affects optimal targeting on heterogeneous networks.
For our experiments, we generate networks following preferential attachment rules with every new node linking to two existing ones (leading to networks with $\langle k\rangle\!\approx4$) and then we allow a random sample of the population to become zealots with uniform zealotry $q_z$, while keeping the remaining nodes as normal agents.

\begin{figure}
    \includegraphics[width=\textwidth]{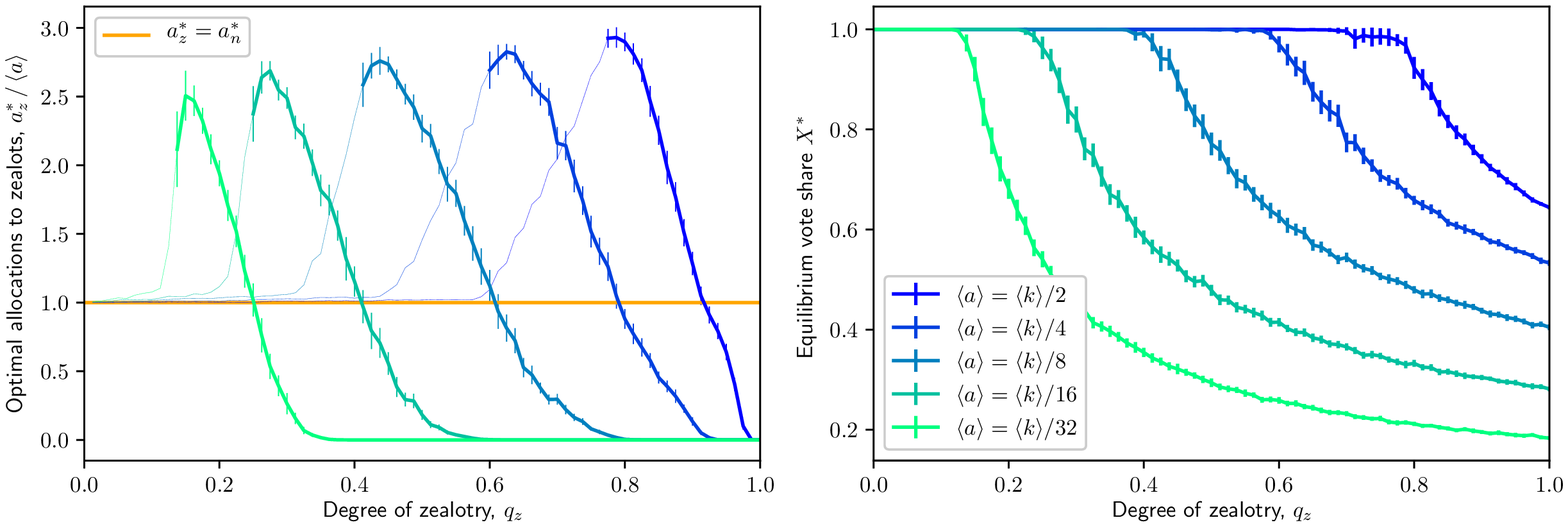}
    \caption{Average optimal allocations given to zealots (left) and resulting equilibrium vote shares (right) on BA networks for varying zealotry $q_z$ and varying per-node budget $\langle a\rangle = \mathcal{B}_a/N$.
    Networks are of size $N=1000$, mean degree $\langle k\rangle\approx4$, and with a random sub-sample of 20\% of the network becoming zealots. Every point is the average over 50 realizations of the experiment, with error bars giving three standard deviations from the mean. Optimal allocations that lead to full control are plotted with a thinner line.
    The horizontal line (\emph{orange}) represents equal link weights given to both groups.} \label{fig:BA_a_and_X}
\end{figure}

We first explore general behaviors by looking at average optimal allocations given to zealots and the resulting equilibrium vote shares (Fig. \ref{fig:BA_a_and_X}). We note remarkable similarities between the results obtained here and those for the complete graph (Fig. \ref{fig:complete}) regarding both optimal allocations and equilibrium vote shares. Note that, while for the complete graph the per-node average allocation is expressed as fractions of nodes in the network ($\langle a\rangle=\mathcal{B}_a/N^2$), we take absolute values for the experiments on BA-networks ($\langle a\rangle= \mathcal{B}_a/N$). This results in a re-scaling of $\langle k\rangle/N$ for exerting a similar influence on BA-networks than on complete networks, as discussed in~\cite{Chinellato2015}.
Again, zealot targeting is preferred when levels of zealotry are low or the available budgets are high. The preference slowly shifts to normal agents as $q_z$ increases, eventually assigning them higher influence links (at $q_z^*$) and even focusing the whole budget on them (at $q^{**}_z$).

\begin{figure}
\includegraphics[width=\textwidth]{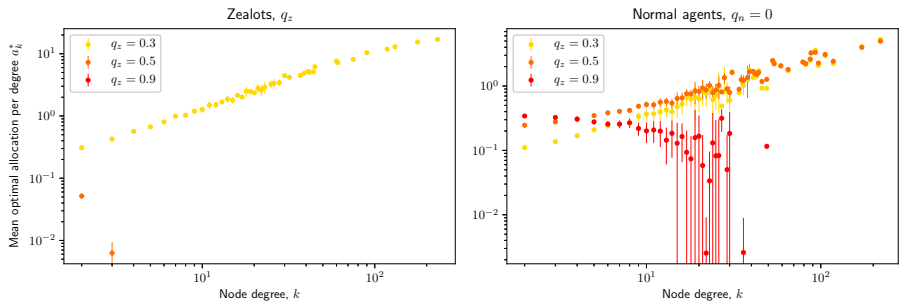}
\caption{Average per-degree allocations given to zealots (left) and normal agents (right) depending on the node degree $k$ and for three values of $q_z$ and budget $\langle a\rangle=\langle k\rangle/16$ on a BA-network size $N=5000$. Each point is the mean allocation over all nodes of same degree, with error bars denoting three standard deviations from the mean.} \label{fig:BA_a_per_degree}
\end{figure}

We next analyze the relationship between optimal allocations and node degree. More specifically, we want to find whether the link preferences to zealots or normal agents are uniformly held across different node degrees.
Figure \ref{fig:BA_a_per_degree} displays optimal allocations $a_k^*$ grouped by node degree $k$ for a given budget $\langle a\rangle=\langle k\rangle/16$ and three different zealotry parameters $q_z=0.3,0.5,0.9$.
We note clear correlations between optimal allocations and node degree in most cases. When the level of zealotry is relatively low ($q_z=0.3$, \emph{yellow}), the external controller exhibits a clear allocation preference to high-degree nodes, among both zealots and normal agents.
For intermediate levels of zealotry ($q_z=0.5$, \emph{orange}) optimal controls omit allocations to high-degree zealots, while mildly target zealots with the lowest degree. This behavior relates to our findings for complete bipartite graphs above, where zealots at the periphery were preferred over hub zealots for low budgets and mild zealotry values. High-degree nodes are still preferred among normal agents in this scenario (right panel) .
Last, when zealots are highly stubborn ($q_z=0.9$, red), they remain untargeted, as well as normal agents on hubs, while the correlation with node degree generally decreases.

\begin{figure}
\includegraphics[width=\textwidth]{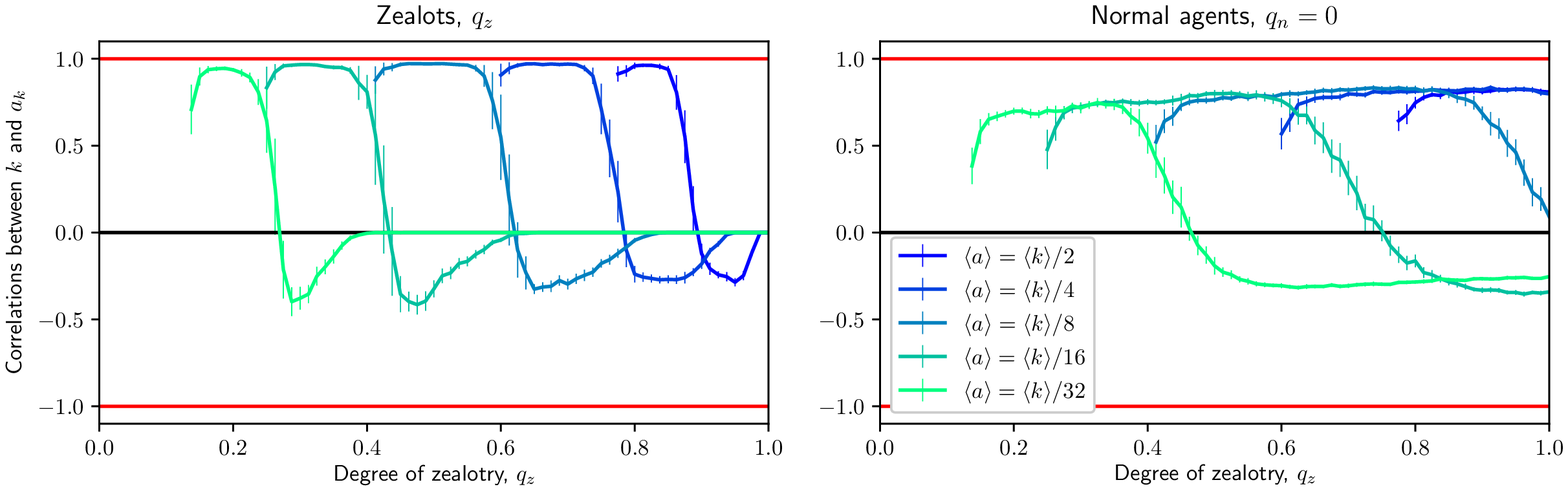}
\caption{Pearson correlations between per-degree optimal allocations $a^*_k$ and node degree $k$ for zealots (left) and normal agents (right) for different values of $\langle a\rangle$ and $q_z$ on BA-networks of size $N=1000$. Each point is the average correlation over 50 runs of the experiment with error bars accounting for three standard deviations from the mean, both computed in the Fisher transformation domain. Correlations are only shown for situations in which full control ($X^*=1$) is not achieved. All p-values fall below $10^{-20}$.} \label{fig:BA_corr}
\end{figure}

We extend the correlation analysis in Fig.~\ref{fig:BA_a_per_degree} to a wider range of zealotry values and available budgets. Figure \ref{fig:BA_corr} shows the Pearson correlations between $a^*_k$ and $k$ on BA-networks for different scenarios. We note again clear patterns that are in agreement with Fig. \ref{fig:BA_a_per_degree}: hubs are preferred when levels of zealotry are low, with the preference decreasing with $q_z$ and switching to periphery nodes at some $q^*_z$. The switching points are lower for zealots than for normal agents and increase with budget availability. Correlations on zealot allocations eventually reach zero, marking the point where they are fully untargeted.

Combining the information from average allocations given to zealots (Fig.~\ref{fig:BA_a_and_X}) and average per-degree allocations given to both groups (Fig.~\ref{fig:BA_corr}), we obtain the following general pattern. When levels of zealotry are low (but not enough for achieving full control of the network), targeting zealots is preferred over targeting normal agents and hubs receive more allocation within each group. Preferences for zealots and zealot hubs diminish as the level of zealotry increases, going through three different transition points. The first transition point $q^*_z$ marks a shift of preference of allocations to normal agents over zealots and preference to zealot with low degree over zealot hubs. After the second transition point~$q^{**}$, no allocation is given to zealots and the preference for hubs among normal agents starts to diminish. After the last transition point $q^{***}_z$, low-degree nodes are preferred over hubs among normal agents.

\section{Conclusions} \label{sect:conclusions}
We have explored influence maximization on heterogeneous networks with zealots biased against the opinion of an external controller. Based on numerical experiments and analytical treatment, a general pattern in optimal influence allocations can be noted. We find that nodes that are harder to control (zealots) receive more influence allocation when the controller's budget is large, while nodes that are easier to control (normal agents) receive more allocation when the budget is small. The transition point between both regimes depends on the level of zealotry and fraction of zealots in the network. When networks with heterogeneous degree distributions are studied, a richer hierarchy emerges, with the following groups ---sorted by their difficulty to be controlled in decreasing order---: zealot hubs, periphery zealots, hub normal agents, and periphery normal agents.
Our findings fit in the general picture of previous literature which has found that optimal allocations tend to depend on a trade-off between budget availability and the difficulty to control nodes \cite{Brede2018,Unpublished,BredeHor2018}.

Although we have uncovered the essential effects that different levels of zealotry have on influence maximization, some additional research questions remain open. For instance, a natural extension to this work could include zealots of both opinions and heterogeneous levels of zealotry in the same network. Similarly, the presence of two opposing external controllers would also be of high interest, as real social influence scenarios tend to include various actors that compete to spread their exclusive opinions. We plan to explore these questions in future work.

\bibliographystyle{splncs04}
\bibliography{bib}

\end{document}